\documentclass[%
 reprint,longbibliography,preprintnumbers,
nofootinbib,
 amsmath,amssymb,
 aps,
prb,
]{revtex4-2}
\pdfoutput=1
\usepackage[utf8]{inputenc}
\usepackage{flushend}
\usepackage{dcolumn}
\usepackage{bm}
\usepackage{balance}

\usepackage[normalem]{ulem}

\usepackage[colorlinks = true,
            linkcolor = blue,
            urlcolor  = blue,
            citecolor =green,
            anchorcolor = blue]{hyperref}
\usepackage{verbatim}
\usepackage{color,ulem}
\usepackage[english]{babel}

\usepackage[utf8]{inputenc}
\input Starburst.fd
\newcommand*\initfamily{\usefont{U}{Starburst}{xl}{n}}\initfamily

\newcommand{\beq}{\begin{eqnarray}}
\newcommand{\eeq}{\end{eqnarray}}
\usepackage{amsmath}
\usepackage{tikz}
\usetikzlibrary{decorations.pathmorphing}
\usetikzlibrary{shapes.misc}
\tikzset{cross/.style={cross out, draw=black, minimum size=8*(#1-\pgflinewidth), inner sep=0pt, outer sep=0pt},
cross/.default={1pt}}
\usetikzlibrary{patterns,math}
\begin{document}

\title{Thickness-dependent conductivity of nanometric semiconductor thin films}

\author{\textbf{Alessio Zaccone}$^{1,2}$}%
 \email{alessio.zaccone@unimi.it}
 
 \vspace{1cm}
 
\affiliation{$^{1}$Department of Physics ``A. Pontremoli'', University of Milan, via Celoria 16,
20133 Milan, Italy.}
\affiliation{$^{2}$Institute for Theoretical Physics, University of G\"ottingen, Friedrich-Hund-Platz 1, 37077 G\"ottingen, Germany}

\begin{abstract}
The miniaturization of electronic devices has led to the prominence, in technological applications, of semiconductor thin films that are only a few nanometers thick. In spite of intense research, the thickness-dependent resistivity or conductivity of semiconductor thin films is not understood at a fundamental physical level. We develop a theory based on quantum confinement which yields the dependence of the concentration of intrinsic carriers on the film thickness. The theory predicts that the resistivity $\rho$, in the 1-10 nm thickness range, increases exponentially as $\rho \sim \exp(const/L^{1/2})$ upon decreasing the film thickness $L$. This law is able to reproduce the remarkable increase in resistivity observed experimentally in Si thin films, whereas the effect of surface scattering (Fuchs-Sondheimer theory) alone cannot explain the data when the film thickness is lower than 10 nm.
\end{abstract}

\maketitle
The physical properties of thin films are pivotal for many technological applications, from optical mirrors to solar cells \cite{solar_science,Friend}, and they are also of fundamental interest for physics. 
Since the advent of transistors \cite{Bardeen}, the Moore's law has consolidated the observation that the number of transistors in a dense integrated circuit (microchip) doubles every two years. This implies that the size of semiconductor blocks in a microchip keeps shrinking as the years go by, and, according to some, it is expected to saturate in the 2020s. Currently, the semiconductor industry is implementing the the ''2 nm process" as the next MOSFET (metal–oxide–semiconductor field-effect transistor) die shrink after the ''3 nm process". While new avenues in microelectronics research are also currently exploring new approaches to more powerful and efficient computing based on neural networks \cite{Science_2024}, the miniaturization of semiconductors at the scale of just few nanometers is still the dominant industrial route for pushing the boundaries of computing power \cite{Irds}.

A quintessential physical property of semiconductors is their electrical conductivity \cite{Sze}. In particular, size effects at the nanometer scale can lead to dramatic changes in the physical properties. In this sense, besides the quest for more powerful computing, miniaturization of semiconductor materials provides a unique opportunity for optimizing and tuning the physical properties to an extent that is impossible to achieve via other modification routes of the bulk material \cite{Grundmann}.

In spite of these tremendous technological implications, the electrical conductivity of nanometer-scale thin films of semiconductors is currently not understood \cite{Duffy}. While it is generally agreed upon that surface effects are no longer discountable when the film thickness approaches few nanometers, a number of fundamental questions arise about whether other effects due to confinement (e.g. quantum confinement effects) may play a role in controlling the thickness dependence of the measured physical properties.

In the following we present evidence, based on the latest experimental data available in the literature, that the observed increase in resistivity by one order of magnitude upon reducing the thickness from 10 nm to 4.5 nm cannot be explained by surface effects (i.e. by the increased carrier scattering at the interface due to increased surface-to-bulk ratio). 
Using a recent quantum confinement theory of thin films, we show that the resistivity due to surface scattering has to be supplemented by the effect of confinement-induced suppression of electronic states, leading to an effective increase of the energy gap and hence to a reduction in intrinsic conductivity. 
The proposed theoretical model is thus able to explain the observed increase by one order of magnitude in resistivity upon approaching 4.5 nm of thickness, which cannot be explained by other models or physical mechanisms.

We schematize the thin film as a 3D material with confinement along the vertical $z$ direction, and we consider it as unconfined along the two other Cartesian directions, i.e. in the $xy$ plane, as schematically depicted in Fig. 1.

\begin{figure}[h]
\centering
\includegraphics[width=\linewidth]{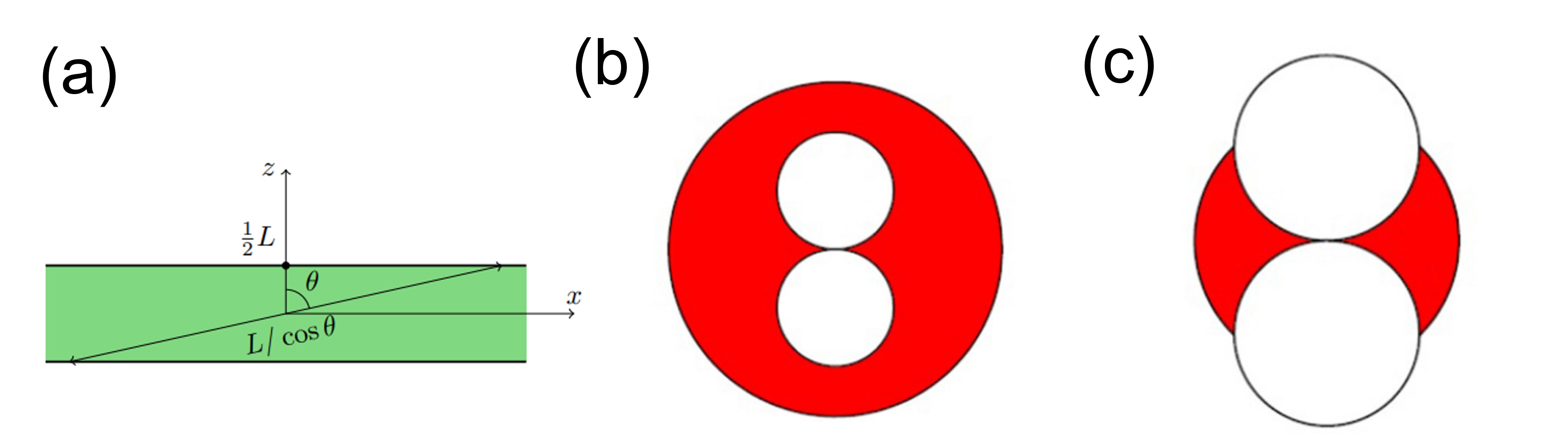}
\caption{Panel (a) shows the thin film geometry in real space (confined along the $z$-axis but unconfined along the $x$ and $y$ axis), with the maximum wavelength of a free carrier that corresponds to a certain polar angle $\theta$. Panel (b) shows the corresponding geometry of electronic states in $k$-space, where the outer Fermi sphere (of radius $k_{F}$) contains two symmetric spheres of hole pockets (states suppressed by confinement), i.e. states in $k$-space that remain unoccupied due to confinement along the $z$-axis. In (c), for strong confinement (e.g. quasi-2D films), the hole pockets of forbidden states have grown to the point that the Fermi surface gets significantly distorted into a surface belonging to a different homotopy group $\mathbb{Z}$. See Refs. \cite{Travaglino_2022,Travaglino_2023,Phillips} for a detailed mathematical derivation of these results. See Refs. \cite{Travaglino_2022,Travaglino_2023,Phillips} for a detailed mathematical derivation of this result. }
\label{fig1}
\end{figure}

As derived in Ref. \cite{Travaglino_2023}, plane-wave electronic states of free carriers with wavelength $\lambda$ larger than $\lambda_{max} = \frac{L}{\cos\theta}$ cannot propagate in the thin film along the direction $\theta$ (cf. Fig. \ref{fig1}(a)), where the polar angle $\theta$ of the propagation direction is measured with respect to the $z$ axis (cf. Fig. \ref{fig1}(a)). 
The momenta $k_{x}$, $k_{y}$, $k_{z}$ are, in general, discretized for small systems if one, arbitrarily, chooses vanishing (hard-wall) boundary conditions (BCs) at the interfaces for the wavefunctions of the governing Schr{\"o}dinger equation. However, if the sample is extended in the $xy$ plane (for at least several microns in both directions), as it is for thin films, $k$ can still be treated as a quasi-continuous variable \cite{Travaglino_2023,valentinis}, because $|\mathbf{k}|=k=2\pi/\lambda$ obeys the following relation \cite{Kittel,Hill}: 
\begin{equation}
\frac{1}{k^{2}}(k_{x}^{2} + k_{y}^{2}+k_{z}^{2})=1.
\end{equation}

There is a further well known reason for treating $k$ and $k_z$ as continuous variables in thin films, in good approximation: it is related to the atomic-scale roughness at the film interfaces, which makes the hard-wall BCs incorrect. It is important to recognize that the hard-wall BCs are a strong idealization of the real physical system, where atomic-scale roughness, disorder and irregularities obviously prevent the wavefunctions to become exactly null at a fixed $z$ coordinate, thus making $k_z$ not a good quantum number. Indeed, it is well known from quantum mechanics, that momentum is a good quantum number for hard-wall or periodic BCs, but not for open BCs. 
As a result of all these facts, it was shown e.g. in Ref. \cite{Yu_2022} (cf. the Supplementary Information therein), that, in real nanometric thin films there is no visible discretization of the wavevector $k_z$ along the confinement direction for non-interacting quasiparticles, even for ultra-thin films with thickness of about 1 nm. 

The confinement-induced cutoff on $\lambda$ means that a number of electronic states in $k$-space are suppressed due to the confinement along the $z$ direction of the film. For a film of thickness $L$, this reduction of the available volume for free carriers in momentum ($k$) space is evaluated exactly as (cf. Fig. \ref{fig1}(b)):
\begin{equation}
    \mathrm{Vol}_{k} = \frac{4}{3} \pi k^3 - 2 \frac{4}{3} \pi \left(\frac{\pi}{L}\right)^3.
\end{equation}

\begin{figure}[h]
\centering
\includegraphics[width=0.9\linewidth]{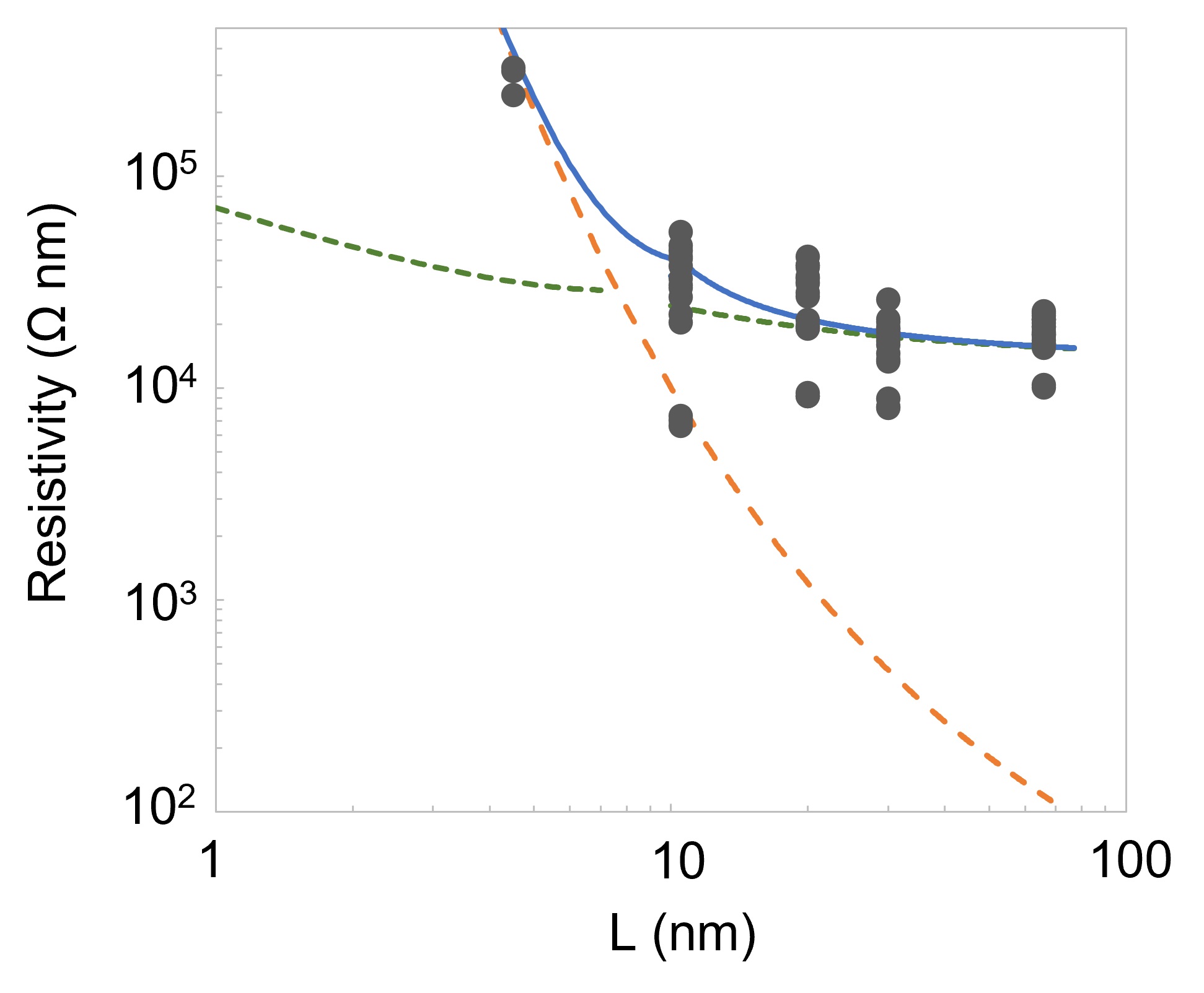}
\caption{Comparison between the theoretical predictions of the proposed model (solid continuous line) obtained by combining the FS surface scattering theory with the electronic confinement model (Eq. \eqref{scaling}) via Matthiessen's rule, Eq. \eqref{Matt}. The green dashed line represents the FS surface scattering prediction (see text for the corresponding equations) without the electron-confinement correction, whereas the orange dashed line represents the electron confinement correction without the FS contribution, given by Eq. \eqref{scaling}. The symbols (circles) are the experimental data from Ref. \cite{Duffy}.  }
\label{fig2}
\end{figure}

Upon reducing the film thickness further below a threshold $L_c=(2 \pi/n)^{1/3}$, where $n$ is the free carrier concentration, one encounters a topological transition described in depth in Ref. \cite{Travaglino_2023}. The Fermi surface gets significantly distorted from the fundamental homotopy group $\pi_1(S^2)=0$ of the spherical surface to a surface belonging to a different homotopy group $\mathbb{Z}$, with the new topology depicted in Fig. \ref{fig1}(c). In this situation, the available volume in $k$ space becomes (see \cite{Travaglino_2023} for a full derivation):
\begin{equation}
    \mathrm{Vol}_k = \frac{4\pi k^3}{3} - V_{inter} = \frac{Lk^4}{2}. 
    \label{case2}
\end{equation}
where $V_{inter}$ denotes the intersection of the two white spheres of hole pockets (states suppressed by confinement) with the original Fermi sphere (Fig. \ref{fig2}). 

From this, the electronic density of states $g(\epsilon)$ of free carriers can be easily evaluated \cite{Travaglino_2023}. It features a crossover from the standard square-root behaviour at higher energy $\epsilon$, to a linear-in-$\epsilon$ behavior at low energy. The crossover is located at an energy $\epsilon^*=\frac{2\pi^2 \hbar^2}{mL^2}$, which depends on the film thickness $L$.

The Fermi level $\mu$ (defined as the Fermi energy $\epsilon_F$ at zero temperature) can then be calculated using standard methods \cite{Travaglino_2023}, leading to:
\begin{equation}
    \mu = \begin{cases} \mu_{\infty} \left(1+\frac{2}{3} \frac{\pi}{n L^3}\right)^{2/3} \mbox{ if } L > L_c = \left(\frac{2\pi}{n}\right)^{1/3} \\ \\
     \frac{(2\pi)^3\hbar^2}{m}\left(\frac{ n}{ L}\right)^{1/2} \mbox{ if } L < L_c = \left(\frac{2\pi}{n}\right)^{1/3}
    \end{cases}\label{chemical}
\end{equation}
where $\mu_{\infty}$ is the Fermi level of the bulk material. 

We consider a c-Si semiconductor thin film, which is either intrinsic or weakly-doped such as e.g. the \emph{ex situ}-doped thin films studied recently in Ref. \cite{Duffy}. Hence, we are going to use equations for the free-carriers concentration as for intrinsic semiconductors. These equations can still be used even for doped semiconductors as long as the concentrations of electrons and holes are comparable \cite{Sze}. Hence, it is assumed that the theoretical model developed below is applicable to thin films doped \emph{ex situ}, which are well known to be affected by poor dopant incorporation and by dopant-deactivation issues, as discussed in Ref. \cite{Duffy}, see also Ref. \cite{Bjork2009}. For nanowires, this effect was explained by simulations in terms of suppressed ionization near the surface due to electronic confinement, with the concomitant formation of image charges outside the wire \cite{Ryu} (cfr. \cite{Weber} for applications of these effects to tailor the conductivity in nanowires). The dopant deactivation effect in ex-situ doped thin films is, however, the result of complex transport phenomena and confinement effects, which cannot be quantitatively modelled in the absence of specific measurements, that are currently out of reach for available experimental techniques.
Also due to this spatially non-uniform dopant-deactivation effect, the doping in these systems is known to be spatially-dependent, with a concentration profile which is generally unknown. Since, anyway, we are interested in the overall resistivity of the sample, which is an averaged quantity over the whole film, we shall emphasize that the carrier concentration in our equations is meant to result from a spatial averaging process over an underlying, as-yet-unknown, spatial profile. While the current model does not consider that all dopants are ionized (on the contrary, given the considerations stated above), and uses an effective, spatially-averaged carrier density, this issue should be kept in mind when considering the comparison with experimental data, in what follows.

For these systems \cite{Duffy}, the concentration of free carriers cannot be determined precisely, and will vary broadly from a lower bound that coincides with the intrinsic material, $n \sim 10^{16}$ m$^{-3}$, to an upper bound of $n \sim 10^{25}$ m$^{-3}$ as noted in \cite{Duffy}.
Here we focus our theoretical analysis on a regime of very weak n-doping where $n \sim 10^{16} - 10^{20}$m$^{-3}$.

In these conditions, $L_c$ is quite large (of the order of hundreds of nanometers). 
Since we are interested in nanometric thin films ($L=1-100$ nm), we can thus safely operate in the regime $L<L_c$, using the second of the two relations reported in Eq. \eqref{chemical}. As it shall become clearer below, we should anticipate that this $L$-dependence due to electron confinement is effectively obscured by the surface-scattering Fuchs-Sondheimer (FS) $L$-dependence down to a certain film thickness at which one observes the crossover from the FS $L$-dependence to the one predicted by electron confinement.

In this regime, the concentration of free carriers can be expressed as \cite{Kittel}:
\begin{equation}
    n_i = \sqrt{n_c(T)n_v(T)}\exp{(-E_g/2k_BT)}
\end{equation}
where $n_c(T)=2 (\frac{m_e^*k_BT}{2 \pi \hbar^2})^{3/2}$ and $n_v(T)=2 (\frac{m_h^*k_BT}{2 \pi \hbar^2})^{3/2}$. Here, $m_e^*$ and $m_h^*$ are the effective masses of electrons and holes, respectively, and $E_g$ is the gap energy.
The latter is related to the Fermi level as follows:
\begin{equation}
    \mu = \frac{1}{2}E_g + \frac{3}{4}k_B T \ln (m_h^*/m_e^*).
\end{equation}
Since holes are lighter than electrons, in general we can write:
\begin{equation}
    E_g = 2 \mu - const \cdot k_B T
\end{equation}
where $const >0$. This relation expresses the well known fact that the Fermi level falls exactly in the middle of the energy gap at $T=0$, while it is shifted upwards towards the bottom of the conduction band at room temperature. It also expresses the fact that the larger the Fermi level, the larger the energy gap. Since, in the thin film, the Fermi level is a function of the film thickness $L$ according to Eq. \eqref{chemical}, the above relation implies that the energy gap $E_g$ is also a function of $L$. In particular, since the Fermi level increases upon decreasing the film thickness $L$, the gap energy $E_g$ will also increase upon decreasing $L$. In the regime $L<L_c$ of relevance for semiconductor thin films, we thus have, for weakly n-doped systems, the following approximate expression for the conductivity $\sigma$:
\begin{equation}
    \sigma = (n_i + n_d) e \mu_e 
\end{equation}
where $n_d$ denotes the concentration of free carriers due to n-doping, e.g. $n_d \approx [n_c(T) N_d]^{1/2} \exp(-E_d/2k_BT)$, where $N_d$ is the concentration of donors and $E_d$ is the ionization energy of the donor impurity atom. As is obvious, nothing in the expression for $n_d$ can ever depend on the film thickness $L$, hence the donor atoms' contribution to the thickness dependence of the conductivity is null. Furthermore, the mobility $\mu_e = e \tau_e/m_e$ is related to the mean free time between collisions of electrons with phonons and impurities, $\tau_e$. Hence, the dependence of the conductivity on $L$ due to the quantum wave confinement of the electrons is given by:
\begin{equation}
    \sigma = (n_i + n_d) e \mu_e \sim \exp{(-const/L^{1/2})}, 
\end{equation}
which is one of the most important results derived in this paper.
The corresponding resistivity contribution due to confinement, $\rho_c$ is then
\begin{equation}
    \rho_c(L) = 1/\sigma \sim \exp{(const/L^{1/2})}. \label{scaling}
\end{equation}
A standard contribution to the resistivity in thin films and other confined materials is due to electron scattering with the interface, or surface scattering \cite{Pop}. This contribution was computed by Fuchs \cite{Fuchs_1938} and Sondheimer \cite{Sond_1952} using the Boltzmann equation framework. As reported by Sondheimer, simple closed-form expressions for the resistivity contribution of surface scattering are obtained for thick films and thin films, respectively, as $\rho_s/\rho_0=[1+3/(8\kappa)]^{-1}$ and $\rho_s/\rho_0=\{4/[3 \kappa \ln (1/\kappa)]\}^{-1}$, with $\kappa=L/\ell$ where $\ell$ is the mean free path in the bulk material \cite{Sond_1952}. For silicon, $\ell \approx 20$ nm. The crossover between the two formulae occurs, therefore, around $10$ nm.
Assuming, as usual, that Matthiessen's rule applies, the different effects of quantum confinement and of surface scattering can be added as independent contributions \cite{Kittel}.
We then have the two thickness-dependent contributions, from quantum wave confinement and from the Fuchs-Sondheimer (FS) scattering, respectively, summing up (Matthiessen's rule) to give the total resistivity $\rho(L)$ as:
\begin{equation}
    \rho(L) = \rho_c(L) + \rho_s(L)\label{Matt}
\end{equation}
where $\rho_s(L)$ is given by the FS theory \cite{Sond_1952}.

Using the form for the confinement-induced resistivity as a function of $L$ given in Eq. \eqref{scaling} for $\rho_c(L)$ and the above quoted asymptotic FS expressions for $\rho_s(L)$, we thus obtain the fitting of the experimental data of Ref. \cite{Duffy} shown in Fig. \ref{fig2}.

As shown in Fig. \ref{fig2}, the dominant contribution to the resistivity is the FS one from surface scattering (green dashed line) down to $L \approx 10$ nm. Up to this value, the confinement-induced contribution to resistivity is orders of magnitude smaller than the FS surface contribution. At about $L \approx 10$ nm we observe a dramatic crossover, whereby the confinement-induced contribution (orange dashed line in Fig. \ref{fig2}) takes over with respect to the FS contribution, and becomes the dominant effect as the thickness is reduced below $10$ nm. Indeed, the confinement-induced contribution to resistivity derived in this paper, Eq. \eqref{scaling}, is able to explain the giant increase of the resistivity (by one order of magnitude) observed in the experimental data upon reducing the film thickness from $10$ nm to $4.5$ nm. It is evident, from Fig. \ref{fig2}, that, without the confinement-induced contribution to resistivity derived here, it would be impossible to capture this sharp increase of the resistivity, with the FS surface-scattering contribution alone.

Regarding the application of the present theory to the experimental data, the following considerations are in order: (i) the theoretical model applies most naturally to the top data points in Fig. 2, which are indeed well fitted by the solid curve (theoretical prediction), since they refer to poorly-doped samples (weakly doped or nearly-intrinsic regime) \cite{Duffy}. These data points exhibit the upturn in resistivity below 10 nm because, indeed, the physics is controlled by electronic confinement, as explained by the model, Eq. \eqref{scaling}, which predicts an exponential increase of resistivity with decreasing $L$ in that range. (ii) the data points that lie at the bottom (i.e. with, comparatively, the lowest resistivity values), as explained in Ref. \cite{Duffy}, correspond to higher active-donors concentrations. Hence, for these data, the mechanism is extrinsic. This consideration is supported by the fact that their resistivity is almost insensitive to the thickness $L$. The proposed theoretical model does not apply to these low-lying data points, but, for completeness, it is important to show them as well as they were reported in Ref. \cite{Duffy}.

Regarding the more general applicability of the proposed model, a few considerations are necessary. The model will be most useful in order to optimize the resistivity, and, hence, the energy and signal transmission efficiency, in materials where the free carrier concentration is low, such that the second of Eq. \eqref{chemical} applies. This is the case of semiconductors (Si, Ge, etc) as well as semi-metals, including quasi-2D semiconductors like molybdenum disulfide and topological semi-metals \cite{Pop}. In these systems, the equations presented above can be used to tune the resistivity of sub-10 nm films by varying the carrier concentration, e.g. by controlling the doping.
For good metals, like Cu, we are, instead, in a regime where the first of Eq. \eqref{chemical} applies, and therefore this effect is smaller and probably negligible with respect to the FS surface-scattering effect \cite{Yuanyue}. Hence, Eq. \eqref{chemical} together with the model presented here, can be used by practitioners to tune and optimize the energetic and signal transmission performance of a given sub-10 nm material of the categories mentioned above, for a certain device.

In summary, we have developed a microscopic theory of confinement-induced conductivity and resistivity in semiconductor thin films which is able to describe and explain the experimental data on c-Si nanometric thin films down to a thickness of about $4.5$ nm. Without the theory presented in this paper, it would be impossible to explain the sharp increase in resistivity, by one order of magnitude, observed experimentally \cite{Duffy} when the film thickness goes below $10$ nm. The theoretical analysis shows that there is a fundamental crossover, at about $10$ nm for the case of weakly-doped c-Si, from a regime (at larger thickness) dominated entirely by the Fuchs-Sondheimer surface-scattering contribution to the resistivity, to a new regime, below $10$ nm, where the resistivity is, instead, entirely dominated by quantum confinement. Interestingly, the same upturn in resistivity with decreasing film thickness below about $10$ nm has been recently observed in first-principles simulations of a very different system such ultrathin copper films \cite{Yuanyue}, which points to a certain degree of universality in the predicted increase of resistivity due to quantum confinement in ultrathin films.
These results will pave the way for a rational understanding and design of new generation microchips with nanometer-scale thickness, within the frame of the "2nm process" \cite{Irds}, and will find broad applications from quantum gated logic and qubits design to solar energy.

\subsection*{Acknowledgments} 
Dr. Ray Duffy (University College Cork and Tyndall National Institute, IE) is gratefully acknowledged for providing the original data set from Ref. \cite{Duffy}.
The author gratefully acknowledges funding from the European Union through Horizon Europe ERC Grant number: 101043968 ``Multimech'', from US Army Research Office through contract nr.   W911NF-22-2-0256, and from the Nieders{\"a}chsische Akademie der Wissenschaften zu G{\"o}ttingen in the frame of the Gauss Professorship program. 

\bibliography{refs}

\begin{thebibliography}{22}%
\makeatletter
\providecommand \@ifxundefined [1]{%
 \@ifx{#1\undefined}
}%
\providecommand \@ifnum [1]{%
 \ifnum #1\expandafter \@firstoftwo
 \else \expandafter \@secondoftwo
 \fi
}%
\providecommand \@ifx [1]{%
 \ifx #1\expandafter \@firstoftwo
 \else \expandafter \@secondoftwo
 \fi
}%
\providecommand \natexlab [1]{#1}%
\providecommand \enquote  [1]{``#1''}%
\providecommand \bibnamefont  [1]{#1}%
\providecommand \bibfnamefont [1]{#1}%
\providecommand \citenamefont [1]{#1}%
\providecommand \href@noop [0]{\@secondoftwo}%
\providecommand \href [0]{\begingroup \@sanitize@url \@href}%
\providecommand \@href[1]{\@@startlink{#1}\@@href}%
\providecommand \@@href[1]{\endgroup#1\@@endlink}%
\providecommand \@sanitize@url [0]{\catcode `\\12\catcode `\$12\catcode `\&12\catcode `\#12\catcode `\^12\catcode `\_12\catcode `\%12\relax}%
\providecommand \@@startlink[1]{}%
\providecommand \@@endlink[0]{}%
\providecommand \url  [0]{\begingroup\@sanitize@url \@url }%
\providecommand \@url [1]{\endgroup\@href {#1}{\urlprefix }}%
\providecommand \urlprefix  [0]{URL }%
\providecommand \Eprint [0]{\href }%
\providecommand \doibase [0]{https://doi.org/}%
\providecommand \selectlanguage [0]{\@gobble}%
\providecommand \bibinfo  [0]{\@secondoftwo}%
\providecommand \bibfield  [0]{\@secondoftwo}%
\providecommand \translation [1]{[#1]}%
\providecommand \BibitemOpen [0]{}%
\providecommand \bibitemStop [0]{}%
\providecommand \bibitemNoStop [0]{.\EOS\space}%
\providecommand \EOS [0]{\spacefactor3000\relax}%
\providecommand \BibitemShut  [1]{\csname bibitem#1\endcsname}%
\let\auto@bib@innerbib\@empty
\bibitem [{\citenamefont {Shah}\ \emph {et~al.}(1999)\citenamefont {Shah}, \citenamefont {Torres}, \citenamefont {Tscharner}, \citenamefont {Wyrsch},\ and\ \citenamefont {Keppner}}]{solar_science}%
  \BibitemOpen
  \bibfield  {author} {\bibinfo {author} {\bibfnamefont {A.}~\bibnamefont {Shah}}, \bibinfo {author} {\bibfnamefont {P.}~\bibnamefont {Torres}}, \bibinfo {author} {\bibfnamefont {R.}~\bibnamefont {Tscharner}}, \bibinfo {author} {\bibfnamefont {N.}~\bibnamefont {Wyrsch}},\ and\ \bibinfo {author} {\bibfnamefont {H.}~\bibnamefont {Keppner}},\ }\bibfield  {title} {\bibinfo {title} {Photovoltaic technology: The case for thin-film solar cells},\ }\href {https://doi.org/10.1126/science.285.5428.692} {\bibfield  {journal} {\bibinfo  {journal} {Science}\ }\textbf {\bibinfo {volume} {285}},\ \bibinfo {pages} {692} (\bibinfo {year} {1999})},\ \Eprint {https://arxiv.org/abs/https://www.science.org/doi/pdf/10.1126/science.285.5428.692} {https://www.science.org/doi/pdf/10.1126/science.285.5428.692} \BibitemShut {NoStop}%
\bibitem [{\citenamefont {Ball}\ \emph {et~al.}(2015)\citenamefont {Ball}, \citenamefont {Stranks}, \citenamefont {Hörantner}, \citenamefont {Hüttner}, \citenamefont {Zhang}, \citenamefont {Crossland}, \citenamefont {Ramirez}, \citenamefont {Riede}, \citenamefont {Johnston}, \citenamefont {Friend},\ and\ \citenamefont {Snaith}}]{Friend}%
  \BibitemOpen
  \bibfield  {author} {\bibinfo {author} {\bibfnamefont {J.~M.}\ \bibnamefont {Ball}}, \bibinfo {author} {\bibfnamefont {S.~D.}\ \bibnamefont {Stranks}}, \bibinfo {author} {\bibfnamefont {M.~T.}\ \bibnamefont {Hörantner}}, \bibinfo {author} {\bibfnamefont {S.}~\bibnamefont {Hüttner}}, \bibinfo {author} {\bibfnamefont {W.}~\bibnamefont {Zhang}}, \bibinfo {author} {\bibfnamefont {E.~J.~W.}\ \bibnamefont {Crossland}}, \bibinfo {author} {\bibfnamefont {I.}~\bibnamefont {Ramirez}}, \bibinfo {author} {\bibfnamefont {M.}~\bibnamefont {Riede}}, \bibinfo {author} {\bibfnamefont {M.~B.}\ \bibnamefont {Johnston}}, \bibinfo {author} {\bibfnamefont {R.~H.}\ \bibnamefont {Friend}},\ and\ \bibinfo {author} {\bibfnamefont {H.~J.}\ \bibnamefont {Snaith}},\ }\bibfield  {title} {\bibinfo {title} {Optical properties and limiting photocurrent of thin-film perovskite solar cells},\ }\href {https://doi.org/10.1039/C4EE03224A} {\bibfield  {journal} {\bibinfo  {journal} {Energy Environ. Sci.}\ }\textbf {\bibinfo {volume} {8}},\
  \bibinfo {pages} {602} (\bibinfo {year} {2015})}\BibitemShut {NoStop}%
\bibitem [{\citenamefont {Bardeen}\ and\ \citenamefont {Brattain}(1948)}]{Bardeen}%
  \BibitemOpen
  \bibfield  {author} {\bibinfo {author} {\bibfnamefont {J.}~\bibnamefont {Bardeen}}\ and\ \bibinfo {author} {\bibfnamefont {W.~H.}\ \bibnamefont {Brattain}},\ }\bibfield  {title} {\bibinfo {title} {The transistor, a semi-conductor triode},\ }\href {https://doi.org/10.1103/PhysRev.74.230} {\bibfield  {journal} {\bibinfo  {journal} {Phys. Rev.}\ }\textbf {\bibinfo {volume} {74}},\ \bibinfo {pages} {230} (\bibinfo {year} {1948})}\BibitemShut {NoStop}%
\bibitem [{\citenamefont {Aimone}\ and\ \citenamefont {Agarwal}(2024)}]{Science_2024}%
  \BibitemOpen
  \bibfield  {author} {\bibinfo {author} {\bibfnamefont {J.~B.}\ \bibnamefont {Aimone}}\ and\ \bibinfo {author} {\bibfnamefont {S.}~\bibnamefont {Agarwal}},\ }\bibfield  {title} {\bibinfo {title} {Overcoming the noise in neural computing},\ }\href {https://doi.org/10.1126/science.adn8545} {\bibfield  {journal} {\bibinfo  {journal} {Science}\ }\textbf {\bibinfo {volume} {383}},\ \bibinfo {pages} {832} (\bibinfo {year} {2024})},\ \Eprint {https://arxiv.org/abs/https://www.science.org/doi/pdf/10.1126/science.adn8545} {https://www.science.org/doi/pdf/10.1126/science.adn8545} \BibitemShut {NoStop}%
\bibitem [{Ird(2022)}]{Irds}%
  \BibitemOpen
  \href {https://irds.ieee.org/editions/2021/more-moore} {\emph {\bibinfo {title} {International Roadmap for Devices and Systems (IRDS™) 2022 Edition}}}\ (\bibinfo  {publisher} {IEEE},\ \bibinfo {year} {2022})\BibitemShut {NoStop}%
\bibitem [{\citenamefont {Sze}\ and\ \citenamefont {Ng}(2006)}]{Sze}%
  \BibitemOpen
  \bibfield  {author} {\bibinfo {author} {\bibfnamefont {S.~M.}\ \bibnamefont {Sze}}\ and\ \bibinfo {author} {\bibfnamefont {K.~K.}\ \bibnamefont {Ng}},\ }\href@noop {} {\emph {\bibinfo {title} {Physics of Semiconductor Devices}}}\ (\bibinfo  {publisher} {John Wiley and Sons, Inc.},\ \bibinfo {year} {2006})\BibitemShut {NoStop}%
\bibitem [{\citenamefont {Grundmann}(2016)}]{Grundmann}%
  \BibitemOpen
  \bibfield  {author} {\bibinfo {author} {\bibfnamefont {M.}~\bibnamefont {Grundmann}},\ }\href@noop {} {\emph {\bibinfo {title} {Physics of Semiconductors}}}\ (\bibinfo  {publisher} {Springer},\ \bibinfo {year} {2016})\BibitemShut {NoStop}%
\bibitem [{\citenamefont {MacHale}\ \emph {et~al.}(2019)\citenamefont {MacHale}, \citenamefont {Meaney}, \citenamefont {Kennedy}, \citenamefont {Eaton}, \citenamefont {Mirabelli}, \citenamefont {White}, \citenamefont {Thomas}, \citenamefont {Pelucchi}, \citenamefont {Petersen}, \citenamefont {Lin}, \citenamefont {Petkov}, \citenamefont {Connolly}, \citenamefont {Hatem}, \citenamefont {Gity}, \citenamefont {Ansari}, \citenamefont {Long},\ and\ \citenamefont {Duffy}}]{Duffy}%
  \BibitemOpen
  \bibfield  {author} {\bibinfo {author} {\bibfnamefont {J.}~\bibnamefont {MacHale}}, \bibinfo {author} {\bibfnamefont {F.}~\bibnamefont {Meaney}}, \bibinfo {author} {\bibfnamefont {N.}~\bibnamefont {Kennedy}}, \bibinfo {author} {\bibfnamefont {L.}~\bibnamefont {Eaton}}, \bibinfo {author} {\bibfnamefont {G.}~\bibnamefont {Mirabelli}}, \bibinfo {author} {\bibfnamefont {M.}~\bibnamefont {White}}, \bibinfo {author} {\bibfnamefont {K.}~\bibnamefont {Thomas}}, \bibinfo {author} {\bibfnamefont {E.}~\bibnamefont {Pelucchi}}, \bibinfo {author} {\bibfnamefont {D.~H.}\ \bibnamefont {Petersen}}, \bibinfo {author} {\bibfnamefont {R.}~\bibnamefont {Lin}}, \bibinfo {author} {\bibfnamefont {N.}~\bibnamefont {Petkov}}, \bibinfo {author} {\bibfnamefont {J.}~\bibnamefont {Connolly}}, \bibinfo {author} {\bibfnamefont {C.}~\bibnamefont {Hatem}}, \bibinfo {author} {\bibfnamefont {F.}~\bibnamefont {Gity}}, \bibinfo {author} {\bibfnamefont {L.}~\bibnamefont {Ansari}}, \bibinfo {author} {\bibfnamefont {B.}~\bibnamefont {Long}},\
  and\ \bibinfo {author} {\bibfnamefont {R.}~\bibnamefont {Duffy}},\ }\bibfield  {title} {\bibinfo {title} {{Exploring conductivity in ex-situ doped Si thin films as thickness approaches 5 nm}},\ }\href {https://doi.org/10.1063/1.5098307} {\bibfield  {journal} {\bibinfo  {journal} {Journal of Applied Physics}\ }\textbf {\bibinfo {volume} {125}},\ \bibinfo {pages} {225709} (\bibinfo {year} {2019})},\ \Eprint {https://arxiv.org/abs/https://pubs.aip.org/aip/jap/article-pdf/doi/10.1063/1.5098307/15231155/225709\_1\_online.pdf} {https://pubs.aip.org/aip/jap/article-pdf/doi/10.1063/1.5098307/15231155/225709\_1\_online.pdf} \BibitemShut {NoStop}%
\bibitem [{\citenamefont {Travaglino}\ and\ \citenamefont {Zaccone}(2022)}]{Travaglino_2022}%
  \BibitemOpen
  \bibfield  {author} {\bibinfo {author} {\bibfnamefont {R.}~\bibnamefont {Travaglino}}\ and\ \bibinfo {author} {\bibfnamefont {A.}~\bibnamefont {Zaccone}},\ }\bibfield  {title} {\bibinfo {title} {Analytical theory of enhanced bose–einstein condensation in thin films},\ }\href {https://doi.org/10.1088/1361-6455/ac5583} {\bibfield  {journal} {\bibinfo  {journal} {Journal of Physics B: Atomic, Molecular and Optical Physics}\ }\textbf {\bibinfo {volume} {55}},\ \bibinfo {pages} {055301} (\bibinfo {year} {2022})}\BibitemShut {NoStop}%
\bibitem [{\citenamefont {Travaglino}\ and\ \citenamefont {Zaccone}(2023)}]{Travaglino_2023}%
  \BibitemOpen
  \bibfield  {author} {\bibinfo {author} {\bibfnamefont {R.}~\bibnamefont {Travaglino}}\ and\ \bibinfo {author} {\bibfnamefont {A.}~\bibnamefont {Zaccone}},\ }\bibfield  {title} {\bibinfo {title} {{Extended analytical BCS theory of superconductivity in thin films}},\ }\href {https://doi.org/10.1063/5.0132820} {\bibfield  {journal} {\bibinfo  {journal} {Journal of Applied Physics}\ }\textbf {\bibinfo {volume} {133}},\ \bibinfo {pages} {033901} (\bibinfo {year} {2023})},\ \Eprint {https://arxiv.org/abs/https://pubs.aip.org/aip/jap/article-pdf/doi/10.1063/5.0132820/16766705/033901\_1\_online.pdf} {https://pubs.aip.org/aip/jap/article-pdf/doi/10.1063/5.0132820/16766705/033901\_1\_online.pdf} \BibitemShut {NoStop}%
\bibitem [{\citenamefont {Phillips}\ \emph {et~al.}(2021)\citenamefont {Phillips}, \citenamefont {Baggioli}, \citenamefont {Sirk}, \citenamefont {Trachenko},\ and\ \citenamefont {Zaccone}}]{Phillips}%
  \BibitemOpen
  \bibfield  {author} {\bibinfo {author} {\bibfnamefont {A.~E.}\ \bibnamefont {Phillips}}, \bibinfo {author} {\bibfnamefont {M.}~\bibnamefont {Baggioli}}, \bibinfo {author} {\bibfnamefont {T.~W.}\ \bibnamefont {Sirk}}, \bibinfo {author} {\bibfnamefont {K.}~\bibnamefont {Trachenko}},\ and\ \bibinfo {author} {\bibfnamefont {A.}~\bibnamefont {Zaccone}},\ }\bibfield  {title} {\bibinfo {title} {Universal ${L}^{\ensuremath{-}3}$ finite-size effects in the viscoelasticity of amorphous systems},\ }\href {https://doi.org/10.1103/PhysRevMaterials.5.035602} {\bibfield  {journal} {\bibinfo  {journal} {Phys. Rev. Mater.}\ }\textbf {\bibinfo {volume} {5}},\ \bibinfo {pages} {035602} (\bibinfo {year} {2021})}\BibitemShut {NoStop}%
\bibitem [{\citenamefont {Valentinis}\ \emph {et~al.}(2016)\citenamefont {Valentinis}, \citenamefont {van~der Marel},\ and\ \citenamefont {Berthod}}]{valentinis}%
  \BibitemOpen
  \bibfield  {author} {\bibinfo {author} {\bibfnamefont {D.}~\bibnamefont {Valentinis}}, \bibinfo {author} {\bibfnamefont {D.}~\bibnamefont {van~der Marel}},\ and\ \bibinfo {author} {\bibfnamefont {C.}~\bibnamefont {Berthod}},\ }\bibfield  {title} {\bibinfo {title} {Rise and fall of shape resonances in thin films of bcs superconductors},\ }\href {https://doi.org/10.1103/PhysRevB.94.054516} {\bibfield  {journal} {\bibinfo  {journal} {Phys. Rev. B}\ }\textbf {\bibinfo {volume} {94}},\ \bibinfo {pages} {054516} (\bibinfo {year} {2016})}\BibitemShut {NoStop}%
\bibitem [{\citenamefont {Kittel}(2005)}]{Kittel}%
  \BibitemOpen
  \bibfield  {author} {\bibinfo {author} {\bibfnamefont {C.}~\bibnamefont {Kittel}},\ }\href@noop {} {\emph {\bibinfo {title} {Introduction to solid state physics}}}\ (\bibinfo  {publisher} {John Wiley and Sons},\ \bibinfo {year} {2005})\BibitemShut {NoStop}%
\bibitem [{\citenamefont {Hill}(1960)}]{Hill}%
  \BibitemOpen
  \bibfield  {author} {\bibinfo {author} {\bibfnamefont {T.~L.}\ \bibnamefont {Hill}},\ }\href@noop {} {\emph {\bibinfo {title} {An Introduction to Statistical Thermodynamics}}}\ (\bibinfo  {publisher} {Addison-Wesley, Reading Massachusetts},\ \bibinfo {year} {1960})\ pp.\ \bibinfo {pages} {493--495}\BibitemShut {NoStop}%
\bibitem [{\citenamefont {Yu}\ \emph {et~al.}(2022)\citenamefont {Yu}, \citenamefont {Yang}, \citenamefont {Baggioli}, \citenamefont {Phillips}, \citenamefont {Zaccone}, \citenamefont {Zhang}, \citenamefont {Kajimoto}, \citenamefont {Nakamura}, \citenamefont {Yu},\ and\ \citenamefont {Hong}}]{Yu_2022}%
  \BibitemOpen
  \bibfield  {author} {\bibinfo {author} {\bibfnamefont {Y.}~\bibnamefont {Yu}}, \bibinfo {author} {\bibfnamefont {C.}~\bibnamefont {Yang}}, \bibinfo {author} {\bibfnamefont {M.}~\bibnamefont {Baggioli}}, \bibinfo {author} {\bibfnamefont {A.~E.}\ \bibnamefont {Phillips}}, \bibinfo {author} {\bibfnamefont {A.}~\bibnamefont {Zaccone}}, \bibinfo {author} {\bibfnamefont {L.}~\bibnamefont {Zhang}}, \bibinfo {author} {\bibfnamefont {R.}~\bibnamefont {Kajimoto}}, \bibinfo {author} {\bibfnamefont {M.}~\bibnamefont {Nakamura}}, \bibinfo {author} {\bibfnamefont {D.}~\bibnamefont {Yu}},\ and\ \bibinfo {author} {\bibfnamefont {L.}~\bibnamefont {Hong}},\ }\bibfield  {title} {\bibinfo {title} {The $\omega$3 scaling of the vibrational density of states in quasi-2d nanoconfined solids},\ }\href {https://doi.org/10.1038/s41467-022-31349-6} {\bibfield  {journal} {\bibinfo  {journal} {Nature Communications}\ }\textbf {\bibinfo {volume} {13}},\ \bibinfo {pages} {3649} (\bibinfo {year} {2022})}\BibitemShut {NoStop}%
\bibitem [{\citenamefont {Bj{\"o}rk}\ \emph {et~al.}(2009)\citenamefont {Bj{\"o}rk}, \citenamefont {Schmid}, \citenamefont {Knoch}, \citenamefont {Riel},\ and\ \citenamefont {Riess}}]{Bjork2009}%
  \BibitemOpen
  \bibfield  {author} {\bibinfo {author} {\bibfnamefont {M.~T.}\ \bibnamefont {Bj{\"o}rk}}, \bibinfo {author} {\bibfnamefont {H.}~\bibnamefont {Schmid}}, \bibinfo {author} {\bibfnamefont {J.}~\bibnamefont {Knoch}}, \bibinfo {author} {\bibfnamefont {H.}~\bibnamefont {Riel}},\ and\ \bibinfo {author} {\bibfnamefont {W.}~\bibnamefont {Riess}},\ }\bibfield  {title} {\bibinfo {title} {Donor deactivation in silicon nanostructures},\ }\href {https://doi.org/10.1038/nnano.2008.400} {\bibfield  {journal} {\bibinfo  {journal} {Nature Nanotechnology}\ }\textbf {\bibinfo {volume} {4}},\ \bibinfo {pages} {103} (\bibinfo {year} {2009})}\BibitemShut {NoStop}%
\bibitem [{\citenamefont {Ryu}\ \emph {et~al.}(2015)\citenamefont {Ryu}, \citenamefont {Kim},\ and\ \citenamefont {Hong}}]{Ryu}%
  \BibitemOpen
  \bibfield  {author} {\bibinfo {author} {\bibfnamefont {H.}~\bibnamefont {Ryu}}, \bibinfo {author} {\bibfnamefont {J.}~\bibnamefont {Kim}},\ and\ \bibinfo {author} {\bibfnamefont {K.-H.}\ \bibnamefont {Hong}},\ }\bibfield  {title} {\bibinfo {title} {Atomistic study on dopant-distributions in realistically sized, highly p-doped si nanowires},\ }\href {https://doi.org/10.1021/nl503770z} {\bibfield  {journal} {\bibinfo  {journal} {Nano Letters}\ }\textbf {\bibinfo {volume} {15}},\ \bibinfo {pages} {450} (\bibinfo {year} {2015})},\ \bibinfo {note} {pMID: 25555203},\ \Eprint {https://arxiv.org/abs/https://doi.org/10.1021/nl503770z} {https://doi.org/10.1021/nl503770z} \BibitemShut {NoStop}%
\bibitem [{\citenamefont {Weber}\ \emph {et~al.}(2014)\citenamefont {Weber}, \citenamefont {Ryu}, \citenamefont {Tan}, \citenamefont {Klimeck},\ and\ \citenamefont {Simmons}}]{Weber}%
  \BibitemOpen
  \bibfield  {author} {\bibinfo {author} {\bibfnamefont {B.}~\bibnamefont {Weber}}, \bibinfo {author} {\bibfnamefont {H.}~\bibnamefont {Ryu}}, \bibinfo {author} {\bibfnamefont {Y.-H.~M.}\ \bibnamefont {Tan}}, \bibinfo {author} {\bibfnamefont {G.}~\bibnamefont {Klimeck}},\ and\ \bibinfo {author} {\bibfnamefont {M.~Y.}\ \bibnamefont {Simmons}},\ }\bibfield  {title} {\bibinfo {title} {Limits to metallic conduction in atomic-scale quasi-one-dimensional silicon wires},\ }\href {https://doi.org/10.1103/PhysRevLett.113.246802} {\bibfield  {journal} {\bibinfo  {journal} {Phys. Rev. Lett.}\ }\textbf {\bibinfo {volume} {113}},\ \bibinfo {pages} {246802} (\bibinfo {year} {2014})}\BibitemShut {NoStop}%
\bibitem [{\citenamefont {Khan}\ \emph {et~al.}(2025)\citenamefont {Khan}, \citenamefont {Ramdas}, \citenamefont {Lindgren}, \citenamefont {Kim}, \citenamefont {Won}, \citenamefont {Wu}, \citenamefont {Saraswat}, \citenamefont {Chen}, \citenamefont {Suzuki}, \citenamefont {da~Jornada}, \citenamefont {Oh},\ and\ \citenamefont {Pop}}]{Pop}%
  \BibitemOpen
  \bibfield  {author} {\bibinfo {author} {\bibfnamefont {A.~I.}\ \bibnamefont {Khan}}, \bibinfo {author} {\bibfnamefont {A.}~\bibnamefont {Ramdas}}, \bibinfo {author} {\bibfnamefont {E.}~\bibnamefont {Lindgren}}, \bibinfo {author} {\bibfnamefont {H.-M.}\ \bibnamefont {Kim}}, \bibinfo {author} {\bibfnamefont {B.}~\bibnamefont {Won}}, \bibinfo {author} {\bibfnamefont {X.}~\bibnamefont {Wu}}, \bibinfo {author} {\bibfnamefont {K.}~\bibnamefont {Saraswat}}, \bibinfo {author} {\bibfnamefont {C.-T.}\ \bibnamefont {Chen}}, \bibinfo {author} {\bibfnamefont {Y.}~\bibnamefont {Suzuki}}, \bibinfo {author} {\bibfnamefont {F.~H.}\ \bibnamefont {da~Jornada}}, \bibinfo {author} {\bibfnamefont {I.-K.}\ \bibnamefont {Oh}},\ and\ \bibinfo {author} {\bibfnamefont {E.}~\bibnamefont {Pop}},\ }\bibfield  {title} {\bibinfo {title} {Surface conduction and reduced electrical resistivity in ultrathin noncrystalline nbp semimetal},\ }\href {https://doi.org/10.1126/science.adq7096} {\bibfield  {journal} {\bibinfo  {journal} {Science}\
  }\textbf {\bibinfo {volume} {387}},\ \bibinfo {pages} {62} (\bibinfo {year} {2025})},\ \Eprint {https://arxiv.org/abs/https://www.science.org/doi/pdf/10.1126/science.adq7096} {https://www.science.org/doi/pdf/10.1126/science.adq7096} \BibitemShut {NoStop}%
\bibitem [{\citenamefont {Fuchs}(1938)}]{Fuchs_1938}%
  \BibitemOpen
  \bibfield  {author} {\bibinfo {author} {\bibfnamefont {K.}~\bibnamefont {Fuchs}},\ }\bibfield  {title} {\bibinfo {title} {The conductivity of thin metallic films according to the electron theory of metals},\ }\href {https://doi.org/10.1017/S0305004100019952} {\bibfield  {journal} {\bibinfo  {journal} {Mathematical Proceedings of the Cambridge Philosophical Society}\ }\textbf {\bibinfo {volume} {34}},\ \bibinfo {pages} {100–108} (\bibinfo {year} {1938})}\BibitemShut {NoStop}%
\bibitem [{\citenamefont {Sondheimer}(1952)}]{Sond_1952}%
  \BibitemOpen
  \bibfield  {author} {\bibinfo {author} {\bibfnamefont {E.}~\bibnamefont {Sondheimer}},\ }\bibfield  {title} {\bibinfo {title} {The mean free path of electrons in metals},\ }\href {https://doi.org/10.1080/00018735200101151} {\bibfield  {journal} {\bibinfo  {journal} {Advances in Physics}\ }\textbf {\bibinfo {volume} {1}},\ \bibinfo {pages} {1} (\bibinfo {year} {1952})},\ \Eprint {https://arxiv.org/abs/https://doi.org/10.1080/00018735200101151} {https://doi.org/10.1080/00018735200101151} \BibitemShut {NoStop}%
\bibitem [{\citenamefont {Zhang}\ and\ \citenamefont {Liu}(2024)}]{Yuanyue}%
  \BibitemOpen
  \bibfield  {author} {\bibinfo {author} {\bibfnamefont {C.}~\bibnamefont {Zhang}}\ and\ \bibinfo {author} {\bibfnamefont {Y.}~\bibnamefont {Liu}},\ }\bibfield  {title} {\bibinfo {title} {Electron-surface scattering from first-principles},\ }\href {https://doi.org/10.1021/acsnano.4c07698} {\bibfield  {journal} {\bibinfo  {journal} {ACS Nano}\ }\textbf {\bibinfo {volume} {18}},\ \bibinfo {pages} {27433} (\bibinfo {year} {2024})},\ \bibinfo {note} {pMID: 39325662},\ \Eprint {https://arxiv.org/abs/https://doi.org/10.1021/acsnano.4c07698} {https://doi.org/10.1021/acsnano.4c07698} \BibitemShut {NoStop}%
\end{thebibliography}%

\end{document}